\documentclass{ws-mpla}
\usepackage[super]{cite}
\usepackage{graphicx}
\usepackage{amsmath}
\usepackage[dvips]{color}
\usepackage{bm}

\RequirePackage{xspace}
\def\epem {\ensuremath{e^+e^-}\xspace}
\newcommand{\gev}{\ensuremath{\mathrm{\,Ge\kern -0.1em V}}\xspace}
\newcommand{\kev}{\ensuremath{\mathrm{\,ke\kern -0.1em V}}\xspace}
\newcommand{\gevcc}{\ensuremath{{\mathrm{\,Ge\kern -0.1em V\!/}c^2}}\xspace}
\newcommand{\tev}{\ensuremath{\mathrm{\,Te\kern -0.1em V}}\xspace}
\def\jpsi     {\ensuremath{{J\mskip -3mu/\mskip -2mu\psi\mskip 2mu}}\xspace}
\def\psitwos  {\ensuremath{\psi{(2S)}}\xspace}
\mathchardef\Upsilon="7107
\def\Y#1S{\ensuremath{\Upsilon{(#1S)}}\xspace}
\def\YOneS  {\Y1S}
\def\YTwoS  {\Y2S}
\def\YThreeS{\Y3S}

\def\km   {\ensuremath{{\rm \,km}}\xspace}

\def\mrad{\ensuremath{\rm \,mrad}\xspace}

\newcommand{\lum} {\ensuremath{\mathcal{L}}\xspace} 
\newcommand{\intlum} {\ensuremath{\mathcal{L}_\mathrm{int}}\xspace} 
\newcommand{\thetac}{\ensuremath{\theta_\mathrm{c}}\xspace} 

\newcommand{\be}{\begin{equation}}
\newcommand{\ee}{\end{equation}}
\newcommand{\bc}{\begin{center}}
\newcommand{\ec}{\end{center}}
\newcommand{\bei}{\begin{itemize}}
\newcommand{\ei}{\end{itemize}}

\begin{document}

\markboth{V.~I.~Telnov}{Monochromatization of \epem colliders with a large crossing angle}

\catchline{}{}{}{}{}

\title{Monochromatization of \epem colliders with a large crossing angle}

\author{V.~I.~Telnov}

\address{Budker Institute of Nuclear Physics, Novosibirsk, 630090, Russia \\
Novosibirsk State University, Novosibirsk, 630090, Russia \\
telnov@inp.nsk.su}

\maketitle


\begin{abstract}
The relative center-of-mass energy spread $\sigma_W/W$ at \epem colliders is ${\cal O}(10^{-3})$, which is much larger than the widths of narrow resonances \jpsi, \psitwos, \YOneS, \YTwoS, \YThreeS mesons, tauonium and some others. It's reduction would significantly increase the resonance production rates and open up great opportunities in the search for new physics. In this paper, we propose a new monochromatization method for colliders with a large crossing angle (which can provide a high luminosity). The contribution of the beam energy spread to $\sigma_W$ is canceled by introducing an appropriate energy--angle correlation at the interaction point; $\sigma_W/W \!\!\sim\!\!$ (0.5--1)$\times 10^{-5}$ appears possible.

\end{abstract}

\ccode{PACS 29.20}

\section{Introduction, necessity of \epem monochromatization, existing method}

The point-like nature of the electron and a narrow energy spread are important advantages of \epem colliders. The energy spread occurs due to synchrotron radiation (SR) in the rings as well as beamstrahlung (BS). The energy spread due to SR depends mainly on the beam energy $E_0$ and magnetic radius of the ring $R$, and only weakly on the specific design of the collider. For uniform rings without damping wigglers $\sigma_E/E \approx 0.86 \times 10^{-3} E [\mathrm{GeV}]/\sqrt{R[\mathrm{m}]}$. The energy spreads for some of the existing and planned \epem rings are given in Table~\ref{Table1}.

\begin{table}[htb]
\tbl{Beam energy spread at circular \epem colliders}
{\begin{tabular}{l c c c c c c c c c }
 & VEPP-2000 & BEPC-II & SuperKEKB & FCC-ee \\[1mm]
 \hline \\[-1.5mm]
$E_0$, \gev & 1 &$ \sim 2$ & 4-7 &  62.5 \\
$2\pi R$,\km & 0.024& 0.24 & 3 & 100 \\
$\sigma_E/E, 10^{-3}$ &$ \sim 0.6 $ &$ \sim 0.5$  & 0.7  & 0.6\! (\!w\!/\!o BS) \\
\end{tabular}
\label{Table1}}
\end{table}

One can see that  the invariant mass spread $\sigma_W/W$=$(1/\sqrt{2})\sigma_E/E\sim$ (0.35--0.5)$\times 10^{-3}$. This spread is much greater than the widths of the narrow \epem resonances \jpsi, \psitwos, {\large $\tau_1$}(tauonium), \YOneS, \YTwoS, \YThreeS and the Higgs boson, see Table~\ref{Table2}. The resonance width $\Gamma$ is the full width at half maximum, so one should compare $\Gamma/m$ and FWHM=2.36 $\sigma_W/W \approx$ (0.8--1.2)$\times 10^{-3}$.
\begin{table}[htb]
\tbl{Width of some narrow \epem resonances}
{\begin{tabular}{l c c c c c c c}
& \jpsi & \psitwos &{\large $\tau_1$}$(\tau^+\tau^-)$  & \YOneS & \YTwoS & \YThreeS & $H(125)$ \\[1mm]
 \hline \\[-2mm]
$m$, \gevcc & 3.097 & 3.686 & 3.554 & 9.460 & 10.023 & 10.355 & 125  \\
$\Gamma$,\kev  & 93& 300 & $2.3\times 10^{-5}$& 54 & 32 & 20.3 & 4200  \\
$\Gamma/m, 10^{-5}$ &3 & 8 & $6.5\times 10^{-7}$& 0.57 & 0.32 & 0.2  & 3.4  \\
$2.36\sigma_W/\Gamma$ &$\sim$35&$\sim$13& $\sim 1.8\times 10^8$&$\sim$180&$\sim$310&$\sim$500&$\sim$30  \\
\end{tabular}
\label{Table2}}
\end{table}

One of the promising directions for particle physics is the study of rare and forbidden processes sensitive to new physics. Therefore, \jpsi and $\Upsilon$ factories with a narrow invariant-mass spread would be good candidates for future experimental facilities.  In the case of a large continuum background, the signal-to-noise ratio $S/\sqrt{B} \propto (\intlum/\sigma_W)/\sqrt{\intlum}=\sqrt{\intlum}/\sigma_W)$; therefore, the integrated luminosity \intlum required to observe a rare decay of a known resonance (or to observe a narrow resonance with a very small $\Gamma_{\epem}$) $\intlum \propto (1/\sigma_W)^2$. A 100-fold improvement in monochromaticity for $\Upsilon$-mesons would be equivalent to a luminosity increase by a factor of $100^2$ = $10000$! In the absence of a background, monochromatization lowers the  branching limit proportionally to $\sigma_W/\intlum$.

The first consideration of energy monochromatization in \epem collisions dates back to mid-1970s~\cite{renieri}. In the proposed scheme, beams collide head-on and have a horizontal or vertical energy dispersion at the interaction point (IP), opposite in sign for the $e^+$ and $e^-$ beams. As a result, the particles collide with opposite energy deviations, $E_0+\Delta E$ and $E_0 -\Delta E$, and their invariant mass $W \approx 2\sqrt{E_1E_2} \approx 2E_0-(\Delta E)^2/E_0$ is very close to $2E_0$. This monochromatization scheme was considered by many authors in 1980s--1990s~\cite{protopopov, avdienko, Wille, Jowett, alexahin, zholents, faus} for use in $c$-$\tau$ and $B$-factory projects (i.e., in the energy range of the $\psi$ and $\Upsilon$ resonances); however, none of the proposals were implemented. The KEKB and PEP-II B-factories operated at a wide \Y4S resonance where monochromatization was not required, high luminosity was more important. Note, this method of monochromatization is associated with an increase of the transverse bunch size ($\sigma_y$ in the case of vertical dispersion), which leads to unacceptable decrease  of the luminosity as $\lum \propto \sigma_W$. This loss of luminosity can only be partially compensated for by a decrease of the horizontal beam size.


\section{New method of monochromatization for colliders with a large crossing angle}

The new generation of circular \epem colliders (DA$\Phi$NE\cite{crab}, SuperKEKB\cite{SuperKEK}, $c$-$\tau$\cite{c-tau,tau-c}, FCC-ee\cite{FCC}, CEPC\cite{CEPC}) rely on the so-called ``crab-waist'' collision scheme~\cite{crab,SuperB}, where the beams collide at an angle $\thetac \gg \sigma_x/\sigma_z$.The maximum luminosity for head-on collisions $\lum\propto 1/\sigma_z$, where $\sigma_z$ is the bunch length, while for collisions at an angle $\lum\propto 1/\beta_y$, where the vertical beta function $\beta_y \sim \sigma_x/\thetac$ can be $\sim 20\text{--}30$ times smaller than $\sigma_z$; as a result, the luminosity can be higher by the same factor. In existing designs, the crossing angle \thetac varies between $30\mrad$ (FCC-ee) and $83\mrad$ (SuperKEKB), it is considered to be $60\mrad$ for $c-\tau$ factories.  In what follows, we propose and explore significant modifications to this collision scheme aimed at achieving monochromatization.
\begin{figure}[!htb]
\centering
\hspace{0.0cm} \includegraphics[width=7.0cm]{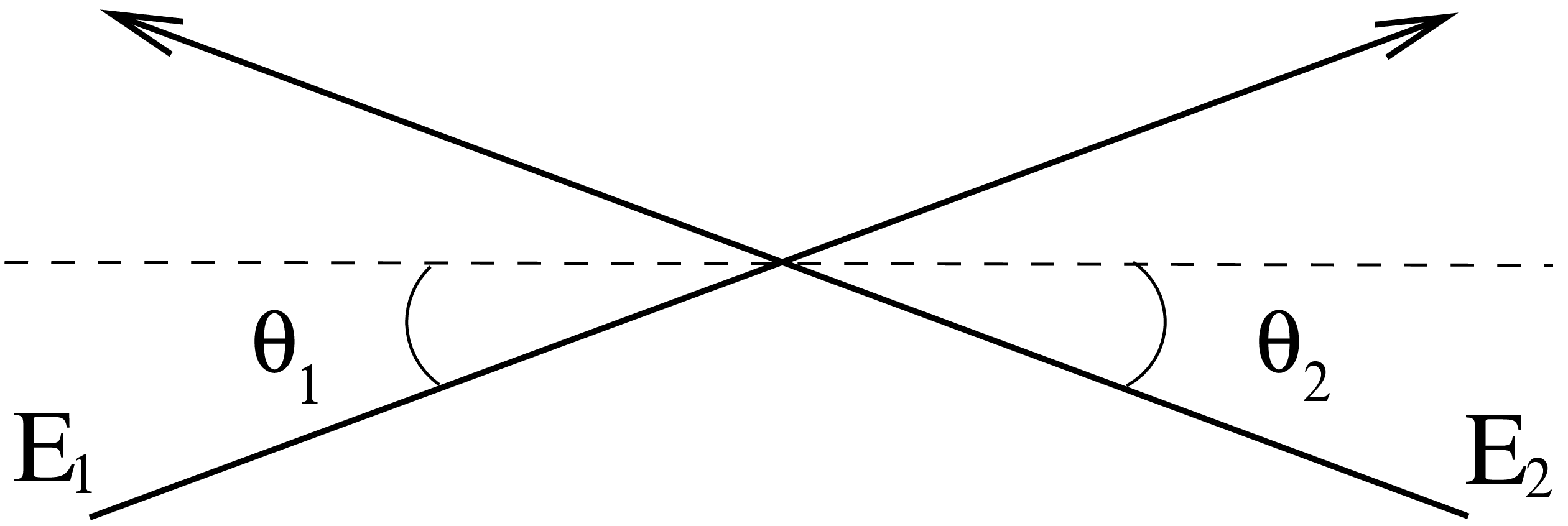}
\vspace{-0.1cm}
\caption{Collisions with crossing angles. }
\label{col-a} \vspace{-0.2cm}
\end{figure}

First, let us consider the mass resolution in the unmodified collision scheme with a crossing angle in the horizontal plane, Fig~\ref{col-a}.
The invariant mass of the produced system ($c=1$)
\be
W^2=(P_1+P_2)^2 \approx 2E_1E_2(1+\cos(\theta_1+\theta_2)).
\label{W2}
\ee
Here, we neglect the terms of the relative order $(m/E)^2$. The contribution of the vertical angular spread is negligible in all practical cases.

By differentiating this formula while assuming that the energies and the angles are independent and setting $\theta_1=\theta_2=\thetac/2$ and $E_1 = E_2 = E_0$, we find the relative mass spread \vspace*{-2mm}
\be
\left(\frac{\sigma_W}{W}\right)^2=\frac{1}{2}\left(\frac{\sigma_E}{E}\right)^2+\frac{1}{2}\frac{\sin^2{\thetac}}{(1+\cos\thetac)^2}\sigma_{\theta}^2,
\label{res0}
\ee
where $\thetac$ is the beam crossing angle, $\sigma_E$ is the beam energy spread, and $\sigma_{\theta}$ is the beam angular spread at the IP, which is determined by the horizontal beam emittance $\varepsilon_x$ and beta function $\beta^*_x$ at the IP: $\sigma_{\theta}=\sqrt{\varepsilon_x/\beta^*_x}$. For head-on collisions, the second term vanishes, and the mass resolution is determined solely by the beam energy spread. In the aforementioned colliders with the crab-waist scheme, the contribution of beam energy spread is also dominant.

The presently proposed monochromatization method is based on the fact that the invariant mass $W^2$ depends on both the beam energies and their crossing angle. The second term in Eq.~\ref{res0} reflects the natural stochastic beam spread due to the horizontal beam emittance and cannot be avoided; however, the first term can be suppressed very significantly, as we shall demonstrate.

In the proposed method, we provide the beams with an angular dispersion $d\theta/dE$ such that a beam particle arrives to the IP with a horizontal angle that depends on its energy: the higher the energy, the larger the angle. By differentiating Eq.~\ref{W2} with the initial values  $\theta_1=\theta_2=\thetac/2$ and $E_1 = E_2 = E$, we find
\be
\mathrm{d}(W^2) = 2E_0\sum_{i=1,2}[(1+\cos\thetac)\mathrm{d}E_i - E_0\sin\thetac \mathrm{d}\theta_i].
\label{dWlin}
\ee
One can see that $\mathrm{d}(W^2)=0$ for the energy-angle correlation (in each beam)
\be
\mathrm{d}\theta_i=\frac{1+\cos\thetac}{\sin\thetac}\frac{\mathrm{d}E_i}{E_0}.
\label{disp}
\ee
Thus, using a certain energy-angle correlation in the colliding beams, we can cancel (in a linear approximation) the contribution of the beam energy spread to the spread in the invariant mass. Note that the proposed monochromatization method works even for unequal beam energies.

Since the first derivative of $W$ is zero, we must use the quadratic term of the Taylor series
\be
\mathrm{d}W^2=(1/2!)\,\mathrm{d}^2W^2(E_1,E_2,\theta_1,\theta_2),
\ee
 where $W^2$ is given by Eq.~\ref{W2} and
\be
\mathrm{d}^2W^2\!=\!\left(\!\mathrm{d}E_1\frac{\partial}{\partial E_1}\!+\!\mathrm{d}E_2\frac{\partial}{\partial E_2}\!+\!\mathrm{d}\theta_1\frac{\partial}{\partial \theta_1}\!+\!\mathrm{d}\theta_2\frac{\partial}{\partial \theta_2}\!\right)^2 \!W^2.
\ee
Then, in the resulting expression we replace $\mathrm{d}\theta_1$ and $\mathrm{d}\theta_2$ by $\mathrm{d}E_1$ and $\mathrm{d}E_2$ using Eq.~\ref{disp}. As a result, we get an expression in the form
$\mathrm{d}W/W=a((\mathrm{d}E_1)^2+(\mathrm{d}E_2)^2))+b(\mathrm{d}E_1 \, \mathrm{d}E_2)$.
In the case of linear dispersion (Eq.~\ref{disp}), both terms contribute and, in addition to fluctuations, there is also a small shift of the mean invariant mass, $\Delta W$.
In the case of Gaussian beam energy distributions with r.m.s.\ spread $\sigma_E$, we have $\overline{(\mathrm{d}E)^2}=\sigma_E^2$, $\sigma(\mathrm{d}E)^2=\sqrt{2}\sigma_E^2$, $\overline{\mathrm{d}E_1\,\mathrm{d}E_2}=0$, $\sigma(\mathrm{d}E_1\,\mathrm{d}E_2)=\sigma_E^2$,\; the mass spreads from the two beams must be summed quadratically. In addition, the fluctuations of the first and the second terms are independent and must be summed quadratically.

Finally, the mass spread due to the beam energy spread and the mass shift are
\be
\small
\left(\frac{\sigma_W}{W}\right)_E \!= \frac{\sigma_E^2}{2E^2}\!\left[\left(1\!+\!\frac{1+\cos{\thetac}}{\sin^2\thetac}\right)^2 \!+\!
\left(\frac{1+\cos{\thetac}}{\sin^2{\thetac}}\right)^2\right]^{1/2}\!\!\!,
\label{we1}
\ee 
\be
\frac{\Delta W}{W}=-\frac{\sigma_E^2}{2E^2}\left(1+\frac{1+\cos{\thetac}}{\sin^2\thetac}\right).
\ee

\begin{figure}[!htb]
\centering
\hspace{0.0cm} \includegraphics[width=9cm]{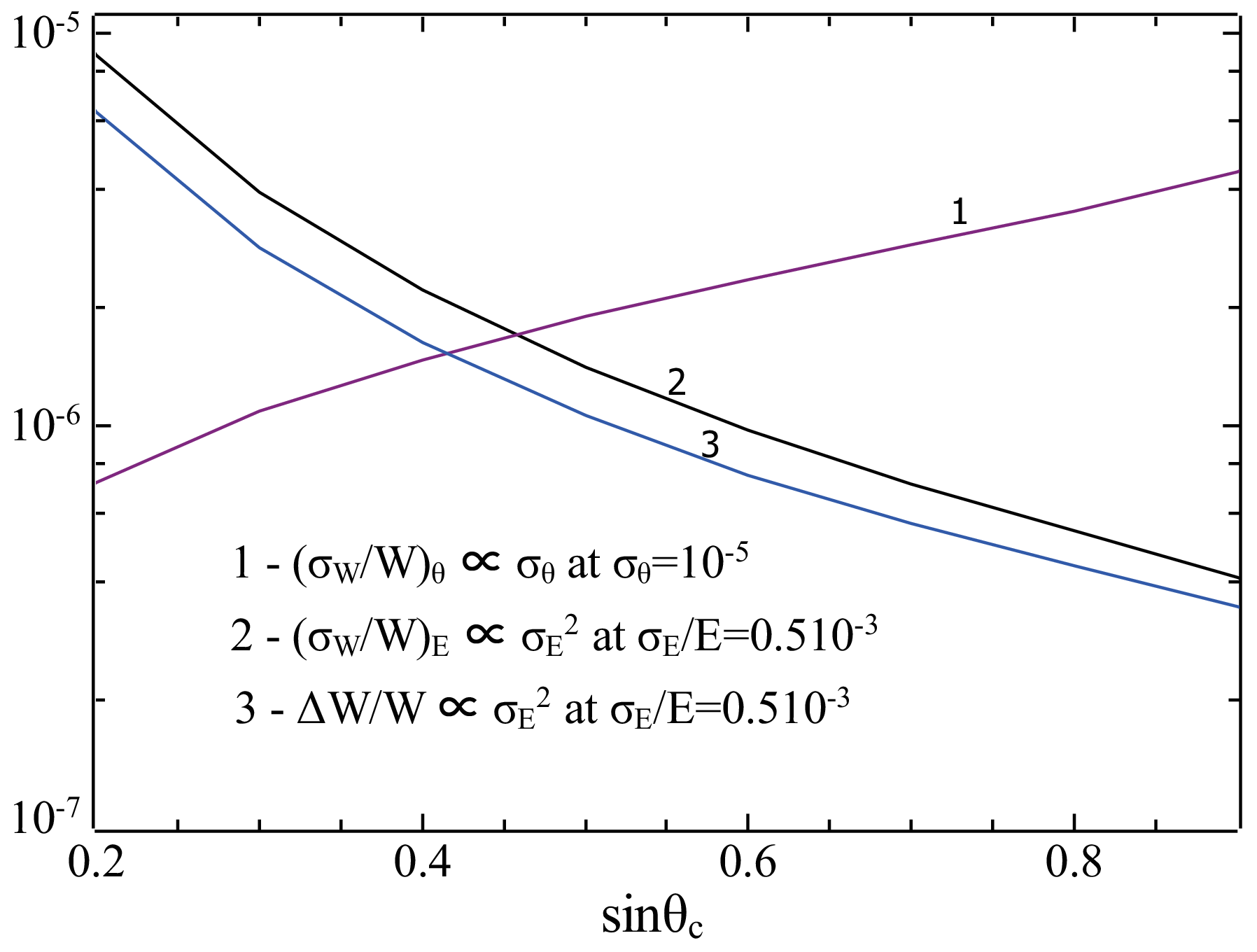}
\caption{Monochromaticity of collisions vs collision angle.}
\label{plot} 
\end{figure}

The total invariant mass spread is the sum of the residual contribution of the energy spread (Eq.~\ref{we1}) and the second term of Eq.~\ref{res0}, which is due to the angular spread: 
\be
\left(\frac{\sigma_W}{W}\right)^2=\left(\frac{\sigma_W}{W}\right)_E^2+\frac{1}{2}\frac{\sin^2{\thetac}}{(1+\cos\thetac)^2}\sigma_{\theta}^2
\label{res}
\ee
These formulas have been verified by direct simulation. The dependence of the invariant mass spread on the collision angle \thetac is shown in Fig.~\ref{plot}, where curves 1 and 2 correspond to contributions of the angular and energy spreads. They should be summed quadratically. The curves are given for $\sigma_{\theta}=10^{-5}$ and $\sigma_E/E=0.5\times 10^{-3}$, these contributions are proportional to $\sigma_{\theta}$ and $\sigma^2_E$.

    It can be seen that the optimal crossing angles lie in the region $\sin{\thetac} \sim$ 0.4--0.5. Let's make a rough estimate of the invariant mass spread by taking $\sin\thetac=0.5$ and the current SCTF parameters: $2E_0=4$ GeV, $\sigma_E/E_0=8.3\times10^{-4}$, $\varepsilon_x=5.8\times 10^{-7}$ cm.  Stochastic beam angular spread at the interaction point (IP) $\sigma_{\theta}=\sqrt{\varepsilon_x/\beta^*_x}$. For $\beta^*_x=1000$ cm we get $\sigma_{\theta}=2.4\times10^{-5}$. Such large $\beta^*_x$ can be used because the beam crossing length $l_c\sim \sigma_x/\sin\thetac=(\varepsilon_x \beta^*_x)^{1/2}/\sin\thetac \approx 0.05$ cm  is small, so you can use similarly small $\beta_y \sim l_c \sim 0.05$ cm, which is good for obtaining high luminosity. Under these assumptions, the contributions of the angular spread and the energy spread are $4.3\times 10^{-6}$ and $3.9\times 10^{-6}$, respectively. After quadratic summation we get $\sigma_W/W\approx 5.8\times 10^{-6}$.

     Realistic figures for achievable monochromatization can only be given after a detailed design has been developed. The main uncertainty relates to the achievable horizontal emittance. Unlike existing projects, the collider under consideration has a region with a high chromatism, which leads to an additional increase of the horizontal emittance due to synchrotron radiation and intra-beam scattering. At the moment, there are only some very preliminary estimates. Very optimistically, one can dream of $\sigma_W/W \sim (0.5-1) \times 10^{-5}$ at $W=$ 3--10 GeV, an improvement up to 50--100 times. That is a very attractive goal!

\section{Possible limitations}
  Below we briefly consider/list some limiting effects.
  \bei
\item  Bunch attraction.
\item  Too strong B in final quadrupoles.
\item  Increase of the horizontal emittance due to emission of synchrotron
radiation and intrabeam scattering in region with a high dispersion
function (final quads, chromatic generation section).
\item Increase of emittances in the detector magnetic field.
\ei
\subsection{Bunch attraction}
At high-luminosity \epem\ factories, the number of particles per bunch is large, $N \approx (5\text{--}10) \times 10^{10}$. The question arises: how does the collision angle change due to the attraction of the beams? Simple estimates indicate that this effect can be problematic. However, a detailed examination unexpectedly shows that beam attraction does not affect the invariant mass of the colliding particles. Indeed, let us consider relativistic particles with the energy $E$ that are attracted by an opposing relativistic beam that creates an electric field $\mathcal{E}$ and a magnetic field $B \approx \mathcal{E}$. At distance $\mathrm{d}s$, the particle receives energy $\mathrm{d}E = e\mathcal{E} \sin{\theta_c}\mathop{\mathrm{d}s} \approx e B \sin{\theta_c} \mathop{\mathrm{d}s}$ and an additional angle $\mathrm{d}\theta \approx (e\mathcal{E}\cos{\thetac}+eB)\,\mathrm{d}s/E \approx eB(1+\cos{\thetac})\,\mathrm{d}s/E$. Substituting $\mathrm{d}E$ and $\mathrm{d}\theta$ for this and similar opposite particle in Eq.~\ref{dWlin}, we show that $\mathrm{d}W=0$!

\subsection{Too strong B in final quadrupoles}
 Due to the required energy/angular dispersion at the IP the horizontal angular spread is (see Eq.\ref{disp})
\be
\sigma_{\theta_x}=\frac{1+\cos\thetac}{\sin\thetac}\frac{\sigma_E}{E_0}.
\label{a-spread}
\ee
For $\sigma_E/E_0=10^{-3}$ and $\sin\thetac =0.5$ we get $\sigma_{\theta_x} \approx 3.7\times 10^{-3}$. The required angular aperture $\theta_x \sim 10\sigma_{\theta_x}\sim 3.7\times 10^{-2}$. The maximum field in the quadrupole can be estimated as $B_{\rm max} \sim E_0\theta_x/eL$. For $E_0=2$ GeV and $L=100$ cm $B_{\rm max} \sim 0.25$ T. So, it is not a problem for energies of $c$-$\tau$ factory and is possibly solvable for the $\Upsilon$ meson region, but not for 100 GeV colliders.

\subsection{Increase of the horizontal emittance due to synchrotron
radiation and intrabeam scattering}
  This method of monochromatization has sections with high chromaticity, this will be lead to an increase of the horizontal emittance due to synchrotron radiation and intrabeam scattering. The reduction of the horizontal emittance is a very important issue, since it determines the achievable monochromatization. To reduce the emittance, it is necessary to use damping wigglers, although they lead to some increase of the energy spread, which contributes to the monochromaticity of collisions. All this needs careful consideration with realistic collider design.

\subsection{Increase of emittances in the detector magnetic field}

The detector is situated in the region with a large dispersion, due to large
crossing angles particles experience a strong magnetic field $B \sim B_s\sin(\thetac/2)$,
which causes the increase of the horizontal and vertical emittance due to synchrotron radiation.
So, solenoidal detector field is almost excluded, one should use the detector without magnetic field in the beam region, toroidal field, for example.

\section{Luminosity with large crossing angle}
A few words about the possible loss of luminosity due to monochromatization. The only difference compared to the SuperKEKB design would be a larger crossing angle, $500\mrad$ instead of $90\mrad$. The luminosity $\lum \propto N(Nf)/(\sigma_z\sigma_y \tan{\thetac/2})$~\cite{SuperB}. For the same beams, an increase of the crossing angle by a factor of 6 would means a loss of luminosity by the same factor. However, the collision effects would become weaker, and one can partially compensate for the loss by increasing $N$ and decreasing $\sigma_z$. The resulting luminosity would be lower, perhaps be a factor of two or three. Such a decrease is acceptable because monochromatization would significantly increase the effective luminosity, $\propto 1/\sigma_W^2$ when studying rare decays in presence of a large background.

\section{Concluding remarks}

In conclusion, a new method of monochromatization of \epem\ collisions is being proposed, which works at large crossing angles ($\thetac \gtrsim $ 0.3--0.5 rad) and allows high luminosities due to reduced collision effects (as in the crab-waist collision scheme). The contribution of the beam energy spread to the invariant mass is compensated by introducing an appropriate energy--angle correlation at the IP. The resulting horizontal angular spread is rather large which limits  applicability of the method to $2E_0 \lesssim 10 \gev$.  The achievable invariant-mass spread is $\sigma_W/W \sim (0.5\text{--}1) \times 10^{-5}$, which is about 50-100 times better than at the past and existing \epem\ storage rings.  Monochromatization is a very natural next step in the development of the next generation of luminosity-frontier colliders. It can increase by several orders of magnitude the effective luminosity in the study of rare decays or looking for narrow states with a small $\Gamma_{\epem}$. The full potential of this method can be realized at the very narrow $\Upsilon(nS)$ resonances as well as at lower energies, where a lot of interesting physics is also present. For example, one can observe and study the tauonium (bound state $\tau^+\tau^-$).  The next step toward realistic projects requires the efforts of accelerator designers.

\section*{Acknowledgments}
The work is supported by the Russian Science Foundation, grant number 24-22-00288.

\end{document}